# Two Dimension Porous Media Reconstruction Using Granular Model Under Influence of Gravity

Pury Sundari[1], Umar Fauzi[2], Zaroh Irayani[2,4] and Sparisoma Viridi[3]

[1]*Physics Study Program, Bandung Institute of Technology, Bandung 40132, Indonesia*
[2]*Physics of Complex Systems, Bandung Institute of Technology, Jl. Ganesha 10, 40132 Bandung, Indonesia*
[3]*Nuclear Physics and Biophysics Research Division, Bandung Institute of Technology, Jl. Ganesha 10, 40132 Bandung, Indonesia*
[4]*Department of Physics, Jenderal Soedirman University, Jl. Dr. Soeparno 61, 53123 Purwokerto, Indonesia*
*purysundari@yahoo.com, umarf@fi.itb.ac.id, zaroh_irayani@yahoo.com, dudung@fi.itb.ac.id*

**Abstract.** Modeling of pores generation in 2-D with granular grains using molecular dynamics method is reported in this work. Grains with certain diameter distribution are let falling due to gravity. Three configurations (larger diameter in upper layer, smaller diameter in upper layer, and mixed) and two kinds of mixture (same grains density and same grains mass) are used in the simulation. Mixture with heterogenous density gives higher porosity than the homogenous one for higher initial height, but change into opposite condition for lower initial height.

## INTRODUCTION

Porous media reconstructions using granular grains are an interesting field nowadays. Using spherical grains as constituent a porous medium can be constructed, where the voids between grains become the pores. Overlaps among grains can be enhanced using hidrostatic compaction [1], random positioning [2], elastic deformation due to grains own weight [3], Poissonian penetrable spheres with radius probability density [4], and deposition of a viscous fluid [5]. Experimental results show the importance of grains geometry that hydraulic conductivity is affected by grains size and form [6]. In this work influence of mass and density accompanied by position configuration of spherical grains are presented and discussed. Molecular dynamics method implementing Euler's method for first order differential equations is used to simulate the spherical grains constructing the porous medium.

## SIMULATION

### Considered Forces

Two spherical grains with diameter $d_i$ and $d_j$ that are positioned at $\vec{r}_i$ and $\vec{r}_j$ will have a repulsive normal force as they are in contact with overlap $\xi_{ij}$

$$\xi_{ij} = \max\left[0, \tfrac{1}{2}(d_i + d_j) - |\vec{r}_i - \vec{r}_j|\right], \quad (1)$$

where $\xi_{ij} = \xi_{ji}$. The simplest model of normal force between grain $i$ and grain $j$ is known as linear spring-dashpot model, which has constant restitution of coefficient and analytical solution [7]. Formulation of the normal force $\vec{N}_{ij}$ is

$$\vec{N}_{ij} = -\left(k_N \xi_{ij} + \gamma_N \dot{\xi}_{ij}\right)\hat{r}_{ij} \quad (2)$$

with $k_N$ and $\gamma_N$ are proportional constant from overlap dan overlap derivation with respect to time $t$. Similar formulation for interaction between spherical grains $i$ and planar surface $k$ (which act as container wall) is defined as $\vec{N}_{ik}$ [8].

Other considered force is due to grains own weight, which is defined as

$$\vec{W}_i = m_i \vec{g} \quad (3)$$

where $\vec{g}$ is earth gravitational acceleration.

### Molecular Dynamics Method

For grain $i$, from the sum of forces acting on it acceleration $\vec{a}_i$ can be calculated using Newton's second law of motion

$$\vec{a}_i = \frac{1}{m_i}\left(\sum_{j \neq i} \vec{N}_{ij} + \sum_k \vec{N}_{ik} + \vec{W}_i\right) \quad (4)$$

Using Euler's method for first order differential equations the new velocity $\vec{v}_i$ and position $\vec{r}_i$ can be found through

$$\vec{v}_i(t + \Delta t) = \vec{v}_i(t) + \vec{a}_i(t)\Delta t \quad (5)$$

and

$$\vec{r}_i(t + \Delta t) = \vec{r}_i(t) + \vec{v}_i(t)\Delta t \quad (6)$$

More complicated method in finding the motions parameter $\vec{v}_i$ and $\vec{r}_i$, such as Gear predictor-corrector algorithm, can also be used as substitution to Euler's method for similar case related to granular grains interaction [9].

## Grains size, mass, and density

There are four different grain diameters which are $0.2\,l_c$, $0.15\,l_c$, $0.1\,l_c$, and $0.075\,l_c$, where $l_c$ is lattice cell. For the homogenous density mixture grains masses are set to 1 g, 0.75 g, 0.5 g, and 0.375 g, for each diameter group, respectively, and for the heterogenous density mixture all grains size are set to have equal mass of 0.5 g.

## Porosity Calculation

Two-dimensional images of spherical grains are converted to black and white image. Suppose that the grains are painted in black, then the voids or pores will be painted in white. Using a code from [2] the fraction of white area divided by whole occupied area is the porosity.

## Pore radius distribution

By considering that a pore have an area $A$ and circumference $C$, the radius of the pore $r_P$ can be determined through

$$r_P = \frac{2A}{C} \quad (5)$$

The pores radius distribution is made by sampling all pores with previous code [2].

## RESULTS AND DISCUSSION

In the simulation following parameters are used: $k_N = 700$, $\gamma_N = 15$, $\Delta t = 0.01$, $y_0 = 0.25\,l_C$, and $0.5\;l_C$. There are three different initial mixture configurations: (a) random position of smaller and larger grains (mixed), (b) larger grains in upper layer (BNE), and (c) smaller grains in upper layer (RBNE). Those three configurations are accompanied by two types of grains density: homogenous and heterogenous. The abbreviation BNE and RBNE are refered to Brazil-nut effect and reverse Brazil-nut effect, respectively [10].

## View of final configurations

Grains final mixture configuration results are given in Figure 1 and 2 for homogenous and heterogenous density, respectively. Both results are used the same three different initial configurations (mixed, BNE, and RBNE). From both figures, the influence of grains density (homogenous or heterogenous) in determining porosity distribution can not be sensed directly.

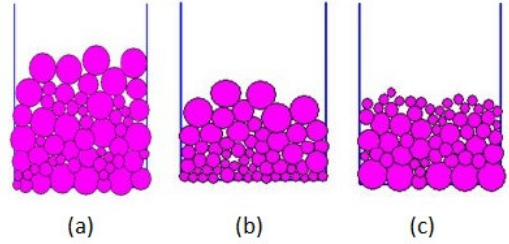

**FIGURE 1.** Final mixture configuration of grains with homogenous density for different initial mixture configuration: (a) mixed, (b) BNE, and (c) RBNE.

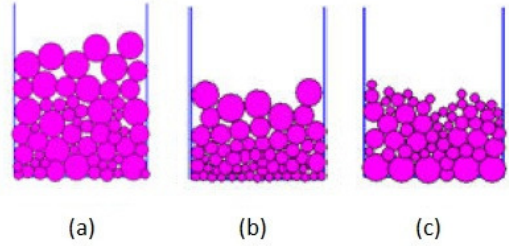

**FIGURE 2.** Final mixture configuration of grains with heterogenous density for different initial mixture configuration: (a) mixed, (b) BNE, and (c) RBNE.

## Porosity

Sampling of results in Figure 1 and 2 using previous code [2] gives the value of porosity for each configuration as given in Table 1.

**TABLE 1.** Porosity range for different initial configurations with $y_0 = 0.5\,l_C$.

| Mixture density | Initial configuration | Porosity range (%) |
|---|---|---|
| Homogenous | Mixed | 9.6 - 9.7 |
| | BNE | 8.8 - 10.7 |
| | RBNE | 6.4 - 7.5 |
| Heterogenous | Mixed | 12.5 - 13.3 |
| | BNE | 10.9 - 14.3 |
| | RBNE | 8.6 - 9.7 |

In overall, heterogenous density gives higher porosity then the homogenous one.

## Pores Radius Distribution

Pore radius distribution for homogenous and heterogenous density mixture are presented in Figure 3 and 4, respectively.

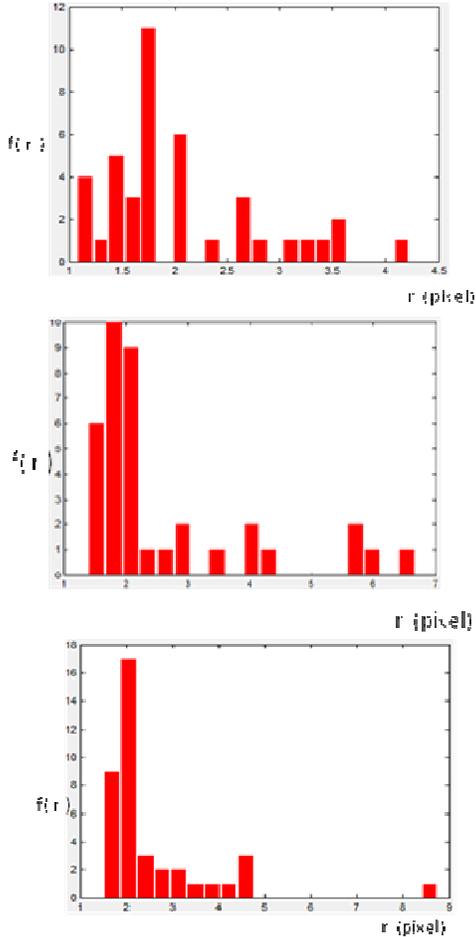

**FIGURE 3.** Pore radius distribution of homogenous grains density for different initial mixture configuration: RBNE (top), BNE (middle), and mixed (bottom).

The top, middle, and bottom of Figure 3 is related to porosity 7.5 %, 10.7 %, and 9.7 %, respectively, while the top, middle, and bottom of Figure 4 is related to porosity 9.7 %, 14.3 %, and 13.3 %.

## Influence of Initial Height

Until now initial height of $y_0 = 0.5\,l_C$ is used in the simulation, which gives results such as Figure 3, Figure 4, and Table 1. If the value of $y_0 = 0.25\,l_C$ is used, then different results are found as in Table 2.

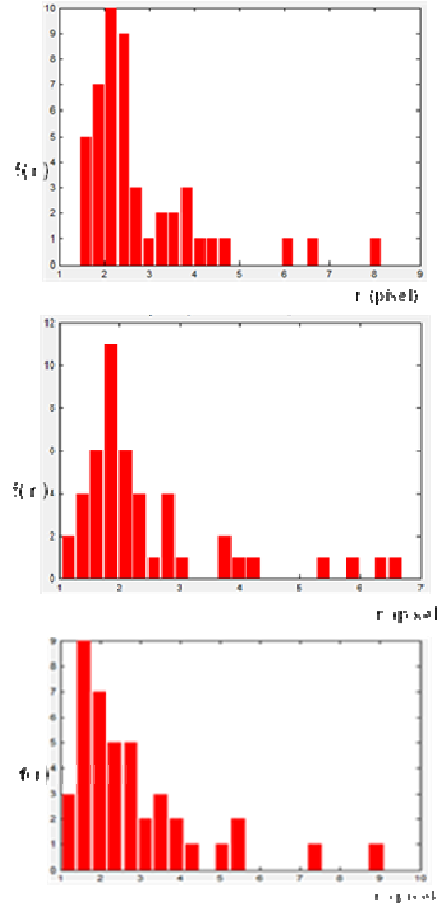

**FIGURE 4.** Pore radius distribution of heterogenous grains density for different initial mixture configuration: RBNE (top), BNE (middle), and mixed (bottom).

Table 1 and Table 2 shows that the initial height $y_0$ has no strong influence in determining the porosity of mixture. Lowering the initial height $y_0$ gives no significant change in porosity range.

**TABLE 2.** Porosity range for different initial configurations with $y_0 = 0.25\,l_C$.

| Mixture density | Initial configuration | Porosity range (%) |
|---|---|---|
| Homogenous | Mixed | 9.2 - 10.2 |
|  | BNE | 11.7 - 13.5 |
|  | RBNE | 5.2 - 6.3 |
| Heterogenous | Mixed | 8.7 - 11.7 |
|  | BNE | 10.3 - 10.7 |
|  | RBNE | 6.7 - 10.8 |

The characteristic of pore radius distribution tends to be log normal distribution. It was obtained by comparing the variance of log normal distibution and normal distibution which are plot on the result of pore radius distribution.

It can be seen that for higher initial height heterogenous density mixture has higher porosity then the homogenous one, but for lower initial height the homogenous density mixture has higher porosity then the heterogenous one. And for lower initial height it is also observed that BNE initial mixture configuration produces highest porosity for both homogenous and heterogenous grains density mixture, which is not found for higher initial height.

Higher porosity in the mixture informs that grains are not well packed, while lower porosity informs otherwise. In the granular field vibrations can rise the density of a granular bed [11], it means that the space between grains become smaller, which in our case is decrease of porosity. For lower initial height RBNE and mixed gives lower porosity, that can be exlained throug the penetration of smaller grains into the space between larger grains. This is why for this initial height BNE initial configuration gives higher porosity. But, why for higher initial height the mixed initial configuration gives highest porosity can be addressed to the experiment that demonstrates how a tennis ball, which is placed on top of a basket ball, jumps higher than its initial height after they fall and hit the ground [12, 13].

## CONCLUSION

Porosity in the mixture of grains is related to the density of the mixture or to the packing of the grains in the mixture. Dropping the grains with several different initial configurations, densities, and initial heights produce different porosity. Detail of mechanisms such as penetration of smaller grains and tennis-ball-on-top-of-basketball effect can explain the results.

## ACKNOWLEGEMENTS

Authors would like to thank to the Director General of Higher Education of Republic of Indonesia for the support through the Hibah Kompetensi grant under the contract no. 227/SP2H/PP/DP2M/III/2010.